\documentclass[nocite]{epl}
\newcommand{\be}{\begin{equation}}
\newcommand{\ee}{\end{equation}}
\newcommand{\ba}{\begin{eqnarray}}
\newcommand{\ea}{\end{eqnarray}}
\usepackage{color}

\usepackage{graphicx}
\usepackage{dcolumn}
\usepackage{bm}
\usepackage{epsfig}

\begin{document}

\pacs{ 81.05.Rm}{Porous material, granular material}
\pacs{ 82.70.-y}{Disperse systems, Complex fluids}
\pacs{ 83.80.Fg}{Granular solids}

\title{Geometric origin of excess low-frequency vibrational modes in weakly-connected amorphous solids.}
\shorttitle{excess modes}
\author{Matthieu Wyart\inst{1}, Sidney R. Nagel\inst{2}, T. A.  Witten\inst{2}}

\institute{\inst{1} Service de Physique de l'Etat Condens\'e (CNRS URA 2464), DSM/DRECAM, CEA Saclay,
91191 Gif sur Yvette, France \inst{2}The James Frank Institute, The University of Chicago, Chicago, Illinois 60637}
\date{\today}

\maketitle

\begin{abstract}

Glasses have an excess number of low-frequency vibrational modes in comparison with most crystalline solids.  We show that such a feature  necessarily occurs in solids with low coordination.  In particular, we analyze the density $D(\omega)$ of normal-mode frequencies $\omega$  and the nature of the low-frequency normal modes of a recently simulated system \cite{J}, comprised of weakly compressed spheres at zero temperature.  We account for the observed a)  convergence of $D(\omega)$ toward a non-zero {\it
constant} as the frequency goes to zero, b) appearance of a low-frequency cutoff $\omega^*$, and c) power-law increase of $\omega^*$ with compression. We introduce  a length scale $l^*$ which characterizes the vibrational modes that appear at $\omega^*$.

\end{abstract}

There is something universal and mystifying about the low-energy behavior of amorphous solids\cite{AndyAnderson,W.Phillips}.  In comparison to most crystals, amorphous solids have a large excess number of low-frequency vibrational modes. In glasses these excitations are seen in the low-temperature specific heat as well as in the spectroscopy of the vibration modes in the terahertz range.  These excitations affect the heat transport \cite{W.Phillips} and might well play an important role at the liquid-glass transition  \cite{parisi} . Nevertheless, little is understood about the cause of such excitations. Whereas in a crystal the vibrations are simply plane waves, in a glass, even at low angular frequency $\omega$, they are much more complicated.   In this Letter, we show that an excess density of vibrational states is a {\it necessary} feature of weakly-connected amorphous solids such as systems with repulsive, short-range interactions. Our analysis elucidates the cause and the peculiar nature of these low-frequency excitations. This gives a new approach for studying some of the ubiquitous phenomena found in glasses. 
 
A dramatic illustration of excess low-frequency vibrations was found in recent computer simulations \cite{ohern, J} of soft-spheres with repulsive, finite range potentials at zero temperature and zero applied shear stress.  These simulations were carried out as a function of the packing fraction, $\phi$ above the jamming threshold, $\phi_c$, where the liquid acquires rigidity and becomes an amorphous solid
\cite{liu}.  O'Hern, Silbert et al. found that the average
number of   contacting neighbors per particle, $z$, the pressure, and the shear modulus vary
as a power of $(\phi - \phi_c)$.  Moreover, these simulations
reveal unexpected features in the density of vibrational mode frequencies, $D(\omega)$: 
(a) As shown in Figure 1,  at $\phi = \phi_c$, when the system is most fragile,
$D(\omega)$ has a plateau extending down to zero frequency with no sign of the
Debye $\omega^2$ density of states normally expected for a three-dimensional
solid.  (b) As shown in the inset to that figure, the plateau erodes
progressively below a frequency $\omega^*$ that increases as $\phi - \phi_c$ increases.  (c)
The value of $D(\omega)$ in the plateau is unaffected by this compression. 
Similar behavior was recently seen in models of tetrahedrally coordinated
covalent glasses \cite{dove}.  Earlier simulations of a Lennard-Jones glass had
also indicated an increase in
$D(\omega)$ at low $\omega$ when $z$ was lowered \cite{Grest}.   

At the jamming transition, $\phi_c$, if particles with no contacts are excluded, the system network of contacts was found to be {\it isostatic}.  That is, there are precisely the right number of inter-particle contacts to prevent free collective motions and the only zero-frequency modes are the uniform translations and rotations of the system as a whole. We shall see that if any contact were to be removed, one
normal mode would become {\it soft}, with $\omega = 0$.  Soft modes have been discussed in relation to various
weakly-connected networks such as covalent glasses \cite{phillips, thorpe},
Alexander's models of soft solids \cite{shlomon}, and models of static forces in
granular packs\cite{Tom1,moukarzel}.  
 
\begin{figure}
\label{fig1}
\centering
\includegraphics[angle=0,width=7cm]{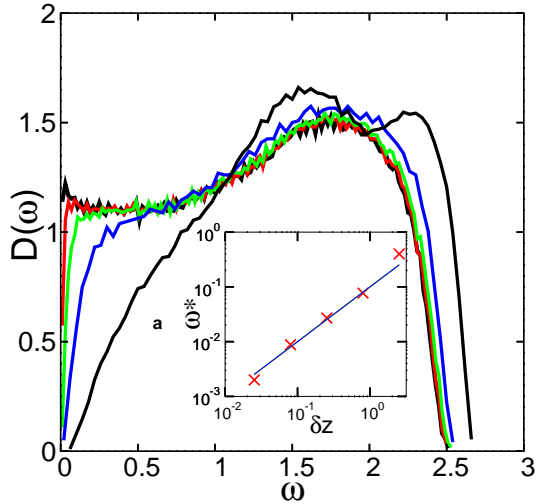}
\caption {$D(\omega)$ {\it vs.} angular frequency $\omega$ for the simulation of Ref \cite{J}.  1024 spheres interacting with repulsive harmonic potentials were compressed in a periodic cubic box to volume fraction $\phi$, slightly above the jamming threshold $\phi_c$.  Then the energy for arbitrary small displacements was calculated and the dynamical matrix inferred. The curve labeled $a$ is at a relative volume fraction $\phi - \phi_c = 0.1$.  Proceeding to the left the curves have relative volume fractions $10^{-2}$,  $10^{-3}$,  $10^{-4}$,  $10^{-8}$, respectively.   Inset: Scaling of $\omega^*$ {\it vs.} $\delta z$.  $\omega^*$ for each $(\phi-\phi_c)$ is determined from the data in the main panel as the frequency where $D(\omega)$ is half of the plateau value. $\delta z$ {\it vs.} $(\phi-\phi_c)$ is obtained from the scaling measured in \cite{J}. The line has slope $1$.
}
\end{figure}

In this paper we explain the plateau of excess low-frequency modes seen in Fig 1 by relating these modes to a certain set of soft modes.  Our procedure for identifying the lowest frequency modes resembles that used for an ordinary solid.  An isolated block of solid has three soft modes that are simply translations along the three co-ordinate axes.  If the  block is enclosed in a rigid container, translation is no longer a soft mode.  However, one may find the lowest-frequency, fundamental modes by making a smooth, sinusoidal distortion of the original soft modes.  We follow an analogous procedure to find the fundamental modes of our isostatic system.  First we identify the soft modes associated with the boundary constraints by removing these constraints.  Next we find a smooth, sinusoidal distortion of these modes that allows us to restore these constraints.

The number of soft modes found in this way is far more than the three fundamental modes of an ordinary solid.  For a d-dimensional cube of side $L$ it is of order $L^{d-1}$.  As in an ordinary solid, the fundamental frequency $\omega_L$ of the distorted modes is of order $L^{-1}$.  The density of these modes $D(\omega_L)$ (per unit volume $L^d$) is thus of order $(L^{d-1}/L^{-1}) L^{-d} \sim L^0$.  In contrast to an ordinary solid, the density remains fixed even as $L \rightarrow \infty$ and $\omega_L \rightarrow 0$.

In the following we make our procedure explicit for the simulation of Ref. \cite{J}.  Using variational bounds, we show that $D(\omega_L)$ cannot go to zero at low frequencies.  We extend our argument to set a lower bound on $D(\omega)$ for an extended range of frequencies.  We account for the effect of compression in suppressing $D(\omega)$ at small $\omega$.  Finally we discuss the applicability of our mechanism to real glasses.

Following \cite{J} we consider $N$ soft spheres packed into a
periodically-continued cube of side $L$ at volume fraction
$\phi$.  For inter-particle distance $r<\sigma$,
the particles are in contact and interact with a potential
$V(r)=\frac{\epsilon} 2 (1-\frac{r}{\sigma})^2$ where $\sigma$ is the particle diameter  and $\epsilon$ a
characteristic energy.  For $r>\sigma$  the potential vanishes and particles do
not interact.  Particles with $r < \sigma$ have non-zero mutual energy and
are said to be in contact. Henceforth we express distance, energy and mass in
units of $\sigma$, $\epsilon$, and $m$, the particle mass,  respectively.   The energy $\delta E$ relative to the equilibrium energy can be
written as a quadratic form in the particle displacements from equilibrium $\delta
\vec{R_1}... \delta
\vec{R_N}$ : 
\ba
\label{1}
\delta E = 	\left [\frac{1}{2}\sum_{\langle  ij \rangle} (r_{ij}^{eq}-1) \frac{[(\delta\vec{R_j}-\delta\vec{R_i})^{\bot}]^2}{2 r_{ij}^{eq}}
\right ] + 
\frac{1}{2}\sum_{\langle  ij \rangle} \left ((\delta\vec{R_i}-\delta\vec{R_j}).\vec n_{ij}\right )^2 +
{\cal O}\{\delta  R^3 \}
\ea 

where the sum is over all $N_c$ contacts $\langle ij\rangle$, $r_{ij}^{eq}$ is the equilibrium distance between particles $i$ and $j$, $\vec n_{ij}$ is the unit vector along the direction $ij$, and $ (\delta\vec{R_j}-\delta\vec{R_i})^{\bot}$ indicates the projection of $\delta\vec{R_j}-\delta\vec{R_i}$ on the plane orthogonal to $\vec n_{ij}$. The term in brackets is proportional to the applied stress, and is much discussed in \cite{shlomon}. In continuous elastic media it is for example responsible for the fast vibrations of strings and drumheads. Near the jamming transition the contact forces vanish, $ r_{ij}^{eq} \rightarrow 1$, and this term becomes arbitrarily small. In this letter we neglect it.

It is convenient to express Eq \ref{1} in matrix form, by defining the set of
displacements $\delta \vec R_1 ... \delta \vec R_N$ as a $3N$-component vector
$|\delta {\bf R}\rangle$.  Then Eq.( \ref{1}) can be written in the form
$\delta E = 
\langle\delta {\bf R}| {\cal M}|\delta {\bf R}\rangle$.  The corresponding
matrix ${\cal M}$ is known as the dynamical matrix\cite{Ashcroft}.  The $3N$
eigenvectors of the dynamical matrix are the normal modes of the particle
system, and its eigenvalues are the squared angular frequencies of these modes.

If the
system does not have enough contacts, $\cal M$ has a set of modes of vanishing
restoring force and thus vanishing vibrational frequency.  For these {\it
soft modes} the energy (the non-bracketed term of Eq.(\ref{1}) in our
approximation) must vanish. Thus the soft modes must satisfy the $N_c$
equations:
\be
\label{2}
(\delta\vec{R_i}-\delta\vec{R_j}).\vec n_{ij}=0 \ \hbox{ for all contacts}\ \langle ij \rangle
\ee
These linear equations define the vector space of displacement fields that conserve the distances to first order between particles in contact. Eq.(\ref{2}) is purely geometrical and does not depend on the interaction potential. Each equation restricts the $3N$-dimensional space of $|\delta{\bf R}\rangle$ by one dimension.  In general, these dimensions are independent, so that the number of independent soft modes is $3N - N_c$. Of these, six modes are dictated by the translational and rotational invariance of the energy function $\delta E[\delta {\bf R}]$. Apart from these, there are $3N - N_c - 6$ independent internal soft modes.

A system of $N$ repulsive particles  is stable if it has no internal soft modes. This leads to the Maxwell criterion \cite{max} for rigidity $N_c \geq
3N-6$.  This is a non-local criterion for stability
against collective motion of particles. At the jamming threshold,
$\phi_c$, this inequality becomes an equality \cite{ohern,Tom1,moukarzel,roux}
and the system is {\it isostatic} with coordination number $z_c\equiv 2N_c/N
\rightarrow 6$, as was found in \cite{J} \footnote{This differs from rigidity
percolation where contacts are deposited randomly on a lattice \cite{yeye}. In
this case when a rigid cluster percolates, it contains over-constrained regions and is not isostatic.}. 

In order to explain the large excess of low frequency modes in isostatic systems, we identify a large set of modes with low frequencies using a variational argument.  Specifically, we shall perturb the boundary conditions in order to reveal a number $p' \sim L^2$ modes whose frequencies are of order
$1/L$.  In order to show the existance
of this many modes, we construct a comparable number of variational trial modes.
 Because ${\cal M}$ is a positive symmetric matrix, if an arbitrary normalized
mode $|\bf \delta R^* \rangle$ has $\delta E=\langle\delta {\bf R^*}| {\cal
M}|\delta {\bf R^*}\rangle \equiv
\omega_t{}^2$
we know that the lowest-frequency eigenmode has a frequency $\omega_0\leq
\omega_t$.  Such an argument can be extended to a set of modes: if there are $m$
{\it orthonormal} trial modes with $\langle\delta {\bf R^*}| {\cal M}|\delta {\bf R^*}\rangle \le
\omega_t{}^2$, then there are at least $m/2$ eigenmodes with frequency
less than $\sqrt 2\omega_t$ \footnote{ If $m_\alpha$ is the $\alpha$'th
lowest eigenvalue of ${\cal M}$ and if $e_\alpha$ is an orthonormal basis such
that $\langle e_\alpha|{\cal M}| e_\alpha\rangle \equiv n_\alpha$ then the
variational bound of A. Horn [Am. J. Math {\bf 76} 620 (1954)] shows that
$\sum_1^p m_\alpha \leq \sum_1^p n_\alpha$.  Since $p n_p \geq \sum_1^p
n_\alpha$, and since $\sum_1^p m_\alpha \geq \sum_{p/2}^p m_\alpha \geq (p/2)
m_{p/2}$, we have $p n_p \geq (p/2) m_{p/2}$ as claimed.} .

\begin{figure}
\label{fig2}
\centering
\includegraphics[angle=0,width=5cm]{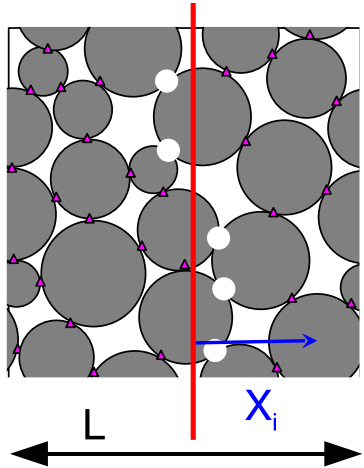}
\caption{Illustration of the boundary contact removal process described in the text.  Eighteen particles are confined in a square box of side $L$ periodically continued horizontally and vertically.  An isostatic packing requires 33 contacts in this two-dimensional system.  An arbitrarily drawn vertical line divides the system.  A contact is removed wherever the line separates the contact from the center of a particle.   Twenty-eight small triangles mark the intact contacts; removed contacts are shown by the five white circles.  }
\end{figure} 

For concreteness we consider the three-dimensional $N$-particle system $\cal S$ of Ref \cite{J} with periodic boundary conditions at the jamming threshold.  We label the axes of the cube by x, y, z.  $\cal S$ is isostatic, so that the removal of $n$  contacts allows exactly $n$ displacement modes with no restoring force.  Consider for example the system $\cal S'$ built from $\cal S$ by removing the $p\sim L^2$ contacts crossing an arbitrary plane orthogonal to (ox) (by convention at $x=0$).  $\cal S'$, which has a free boundary condition instead of periodic ones along (ox), contains a space of soft modes of dimension $p$.  (FIg. 2 illustrates how contacts are removed by changing periodic boundary conditions to free ones in a two-dimensional system.)  In general when a contact $\langle ij \rangle$  is cut in an isostatic system, the corresponding soft mode is not localized near  $\langle ij \rangle$, but extends over the entire system.  This extended character comes from the non-locality of the isostatic condition that gives rise to the soft modes. This was confirmed in isostatic simulations\cite{Tom1}, which observed that the amplitudes of the soft modes were spread over a non-zero fraction of the particles.  Below we assume that a vector space encompassing a non-zero fraction of the soft modes have this extended character.  We then consider the possibility of non-extended modes.

We now use the vector space of dimension $p'\sim L^2$ of extended soft modes of $\cal{S'}$ to build $p'$ orthonormal trial modes of $\cal{S}$ of frequency of the order $1/L$.  Let us define $|\bf \delta R_\beta \rangle$ to be a normalized basis of this space, $1\leq \beta \leq p'$.  These modes are not soft in the jammed system $\cal S$ since they deform the previous $p$ contacts located near $x=0$.  Nevertheless a set of trial modes, $|\bf \delta R_\beta^* \rangle$, can still be formed by altering the soft modes so that they do not have an appreciable amplitude at the boundary where the contacts were severed.  We seek to alter the soft mode to minimize the distortion at the severed contacts while minimizing the distortion elsewhere. Accordingly, for each soft mode $\beta$ we define the corresponding trial-mode displacement $\langle i|{\bf \delta R}^* \rangle$ to be: 
\be
\label{rr}
\langle i|{\bf \delta R}_\beta^* \rangle \equiv C_\beta \sin(\frac{x_i \pi}{L}) \langle i|\bf \delta R_\beta \rangle
\ee
where the constants $C_\beta $  are introduced to normalize the modes. The extended character of the soft modes implies that the $C_\beta$ are of order one, and therefore bounded                              above by a constant $c$ independent of $L$ and $\beta$ \footnote{A normalized mode $\beta$ is said to be extended if there is a constant  $c_2$ independent of N such that $(\langle i|{\bf \delta R}_\beta \rangle)^2> c_2/N$ for a non-vanishing fraction of the particles $i$. It is then straightforward to show  that $C_\beta$ defined as $C_\beta^{-2}\equiv \sum_{\langle ij \rangle} \sin^2(\frac{z_i \pi}{L})   \langle j|\bf \delta R_\beta \rangle^2$ is bounded below by a positive constant $c_1$ independent of N.}. The sine factor suppresses the gaps and overlaps at the  $p$ contacts near $x=0$ and $x=L$. Formally, the modulation by a sine is a linear mapping. This mapping is invertible if it is restricted to the extended soft modes. Consequently the basis $|\bf \delta R_\beta \rangle$ can always be chosen such that the  $|\bf \delta R_\beta^* \rangle$ are orthogonal. This procedure generates a $\delta E$ for each trial mode of order $1/L^2$.  Indeed using Eq.(\ref{1}) and Eq.(\ref{2}), one obtains:

\ba
\label{kk}
\delta E \approx  C_\beta^2 \sum_{\langle ij \rangle} \cos^2(\frac{x_i \pi}{L}) \frac{\pi^2}{L^2} (\vec n_{ij} \cdot \vec e_x)^2 ( \langle j|{\bf \delta R}_\beta \rangle \cdot \vec n_{ij})^2\\
\label{kkk}
\leq   c^2  (\pi/L)^2 z_{max} \equiv \omega_L^2
\ea
where $\vec e_x$ is the unit vector along (ox), and $z_{max}$ is the maximum coordination number for one particle, $z_{max}=12$ for spheres \footnote{ In a polydisperse system $z_{max}$ could a priori be larger. Nevertheless Eq.(\ref{kk}) is a sum on every contact where the displacement of only one of the two particles appears in each term of the sum. The corresponding particle can be chosen arbitrarily. It is convenient to choose the smallest particle of each contact. Thus when this sum on every contact is written as a sum on every particle to obtain Eq.(\ref{kkk}), the constant $z_{max}$ still corresponds to the monodisperse case, as a particle cannot have more contacts with particles larger than itself. }. Thus, we have accomplished our goal of finding $\sim L^2$ trial orthonormal modes below a frequency  $\omega_L\sim 1/L$.  We can now apply the above mentioned variational argument to show that the average density of states is bounded by a constant below frequencies of order $\omega_L$.  

One may ask if the present variational argument can be improved, for example by considering geometries of broken contacts different from the surface we considered up to now.  When contacts are cut to create a vector space of  extended soft modes, the soft modes must be modulated with a function that vanishes where the contacts are broken in order to obtain trial modes of low energy. On the one hand, cutting many contacts increases the number of trial modes. On the other hand, if too many contacts are broken, the modulating function must have many ``nodes'' where it vanishes. Consequently this function displays larger gradients and the energies of the trial modes increase. Cutting a surface appears to be the best compromise between these two opposing effects. Thus our argument gives a natural limit to the number of low-frequency states to be expected.

We may extend this argument to show that the bound on the average density of states extends to a non-zero fraction of the modes of the system.  If the cubic simulation box were divided  into $m^3$ sub-cubes of size $L/m$, each sub-cube must have a density of states equal to the same $\langle D(\omega)\rangle$ as was derived above, but extending to frequencies of order $m\omega_L$.  These subsystem modes must be present in the full system as well, therefore the  bound on $D(\omega)$ extends to $[0,m \omega_L]$.  We thus prove that the same bound on the average density of states holds down to sizes of the order of a few particles, corresponding to frequencies independent of $L$.  We note that in $d$ dimensions this argument may be repeated to yield a total number of modes, $L^{d-1}$, below a frequency $\omega_L \approx 1/L$, thus yielding a limiting non-zero density of states in any dimension.

We now consider how these states are altered when the system is compressed to volume fractions above $\phi_c$. 
The simulations show that the extra-coordination number  $\delta z \equiv (z-z_c) \propto (\phi-\phi_c)^{0.5}$\cite{durian,J}.    Compression causes $\Delta N_c=N \delta z/2 \sim L^3 \delta z$ extra constraints to appear in Eq.(\ref{2}).  Cutting the boundaries of the system, as we did above, relaxes $p\sim L^2$ constraints. For a large system $L^3\delta z> L^2$, thus  $p<\Delta N_c$ and Eq.(\ref{2}) is still over-constrained so that no soft modes appear in the system.  
However, as the systems become smaller, the excess of $\Delta N_c$ diminishes, and for $L$ smaller than some $l^*\sim (\delta z)^{-1}$, $p$ becomes larger than $\Delta N_c$ and  the system is again under-constrained.  This allows one to build the trial modes described above in
subsystems smaller than $l^*$.  These modes appear above a cut-off frequency
$\omega^*\sim (l^*)^{-1}$; they are modes that contribute to the  plateau in
$D(\omega)$ above
$\omega^*$. 
In other words, modes with characteristic length smaller than
$l^*$ are  little affected by the extra contacts, and the density of states is
unperturbed above a frequency $\omega^* \sim \delta z$ \cite {otherlength}. Using the explict potential $V(r)$ defined above Eq.(\ref{1})  $\omega^*\simeq (\epsilon/m)^{1/2}\sigma^{-1}\delta z$.
This scaling is checked numerically in the inset of Fig.1. It is in very good
agreement with our prediction up to $\delta z\approx 2$.

These arguments account for the occurrence of an excess in the low-frequency spectrum of $D(\omega)$ in amorphous systems of repulsive particles with finite-range interactions. The corresponding low-frequency modes are related to the soft modes of free-boundary subsystems of size smaller than $l^*$.  $l^*$  does not appear in the static structure of these amorphous solids, but only in their responses. $l^*$  depends on the {\it connectivity} $z$ and diverges at the jamming transition.  At length scales larger than $l^*$, we expect the system to behave  as a continuous body.

The argument above presumed that one could find a substantial subspace of $p'$  {\it extended} modes in the space of $p$ soft modes created by breaking the boundary contacts.  This presumption, while plausible, is not trivial.  Indeed, one can readily invent ways of breaking contacts that lead to most of the soft modes being localized near the broken contacts rather than being extended.  For example if one breaks all the contacts in a spherical region, most of the spheres become unconstrained, and most of the modes are just the local motions of individual spheres.  More generally compact arrangements of broken contacts lead to localized soft modes. We have checked numerically that the soft modes defined above Eq.(\ref{rr}) are extended in the sense needed \cite{m1}.

The approximate low-frequency modes defined by Eq.(\ref{rr}) may help explain several puzzling properties of glasses. For example, their transport properties are affected at temperatures that correspond to the frequency where the excess of vibrations appears \cite{W.Phillips}.  Our analysis can give deeper insight into the vibrational modes responsible for e.g., thermal conductivity. Moreover, our picture of these modes  suggests a novel mechanism for mechanical and thermal relaxation. They give an alternative to the conventional cage picture in glasses: the non-locality of the soft modes suggests relaxation via extended events, such as the nonlinear buckling of modes resembling those of Eq.(\ref{rr}),  rather than by local jumps out of a cage. The non-locality is a fundamental consequence of the Maxwell criterion for rigidity that underlies our argument.

Finally we discuss the generality of this description of low-frequency vibrations of amorphous solids. On the one hand, because our argument depends only on the connectivity of the system, it should be equally valid for systems with three-body forces (e.g. the model of SiO$_2$ of \cite{dove} which leads to density of states similar to Fig.1) or particles with friction where degrees of freedom of rotation matter, provided the co-ordination number $z$ is appropriately small.  On the other hand, our approach neglects forces between
non-adjacent particles, such as the Van der Waals forces present in all
molecular glasses.  Since these non-adjacent contacts are weak, our approach may
be instructive in these real systems as well.

We  thank Leo Silbert for providing the numerical data and for discussions regarding the simulation results, and J.P Bouchaud, Bulbul Chakraborty, L.E. Chayes, P.G. de Gennes, Andrea Liu and Corey O'Hern for helpful discussions. We also acknowledge the support of the CFR fellowship, MRSEC DMR-0213745 and DOE grant DE-FG02-03ER46088.

\end{document}